\documentclass{article}

\usepackage{arxiv}

\usepackage[utf8]{inputenc} 
\usepackage[T1]{fontenc}    
\usepackage{hyperref}       
\usepackage{url}            
\usepackage{booktabs}       
\usepackage{amsfonts}       
\usepackage{amsmath}
\usepackage{nicefrac}       
\usepackage{microtype}      
\usepackage{cleveref}       
\usepackage{lipsum}         
\usepackage{graphicx}
\usepackage{subcaption}

\usepackage[authoryear,round]{natbib}
\let\cite\citep

\usepackage{doi}
\usepackage{xcolor}
\usepackage{multirow}
\usepackage[most]{tcolorbox}
\usepackage{float}

\newtcolorbox{solutionbox}{
  enhanced,
  breakable,
  colback=gray!8,
  colframe=gray!35,
  boxrule=0.5pt,
  arc=2mm,
  left=2mm,
  right=2mm,
  top=2mm,
  bottom=2mm,
  fontupper=\ttfamily\small
}

\title{Students using GenAI lag behind in problem-solving competence: an agent-based study of classroom networks}

\newcommand{\orcidicon}[1]{%
  \href{#1}{\includegraphics[scale=0.06]{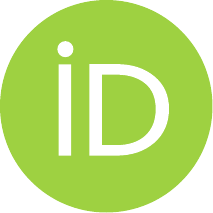}}\hspace{1mm}%
}

\author{%
  \orcidicon{https://orcid.org/0000-0000-0000-0000}Lorenzo Betti\textsuperscript{1}
  \And
  \orcidicon{https://orcid.org/0009-0009-5929-5820}Iacopo Caporossi\textsuperscript{2}
  \And
  \orcidicon{https://orcid.org/0009-0000-3859-0825}Carsten Källner\textsuperscript{3}
  \AND
  \orcidicon{https://orcid.org/0000-0002-1220-0718}Karolina Levanaitė\textsuperscript{4}
  \And
  \orcidicon{https://orcid.org/0000-0002-1297-2160}Chenyu Li\textsuperscript{5}
  \And
  \orcidicon{https://orcid.org/0009-0003-2481-1721}Xuan-Chen Liu\textsuperscript{6}
  \And
  Giulia Lorenzini\textsuperscript{7}
  \And
  \orcidicon{https://orcid.org/0009-0000-7968-7885}Vittoria Socci\textsuperscript{2}
  \AND
  \orcidicon{https://orcid.org/0000-0002-1074-0411}Michele Re Fiorentin\textsuperscript{8}
  \And
  \orcidicon{https://orcid.org/0009-0004-0382-9807}Ilaria Stanzani\textsuperscript{9}
  \And
  \orcidicon{https://orcid.org/0009-0008-9894-4857}Marta Baratto\textsuperscript{7,\textdagger}
  \AND
  \vspace{-.3cm}\\
  {\normalfont\small
  \parbox{0.98\textwidth}{\centering
    \textsuperscript{1}Department of Network and Data Science, Central European University, Vienna, Austria\\
    \textsuperscript{2}Dipartimento di Ingegneria dell'Informazione e Scienze Matematiche, University of Siena, Siena, Italy\\
    \textsuperscript{3}Complexity Science Hub, Vienna, Austria\\
    \textsuperscript{4}Institute of Educational Sciences, Vilnius University, Vilnius, Lithuania\\
    \textsuperscript{5}Department of Computer Science, Aalto University, Espoo, Finland\\
    \textsuperscript{6}Network Science Institute, Northeastern University London, London, United Kingdom\\
    \textsuperscript{7}Department of Physics, University of Turin, Turin, Italy\\
    \textsuperscript{8}Department of Applied Science and Technology, Politecnico di Torino, Turin, Italy\\
    \textsuperscript{9}Dipartimento di Informatica, Bioingegneria, Robotica e Ingegneria dei Sistemi, University of Genova, Genova, Italy\\
    \vspace{.1cm}
    \textsuperscript{\textdagger}{\ttfamily marta.baratto@unito.it}
  }}
}

\hypersetup{
pdftitle={Will everyone become stupid or super-intelligent? The Impact of AI on Education},
pdfsubject={q-bio.NC, q-bio.QM},
pdfauthor={Lorenzo Betti, Iacopo Caporossi, Carsten Källner, Karolina Levanaitė, Chenyu Li, Giulia Lorenzini, Vittoria Socci, Xuan-Chen Liu, Ilaria Stanzani, Michele Re Fiorentin, Marta Baratto},
pdfkeywords={Artificial Intelligence, Education, Collective Learning},
}

\begin{document}
\maketitle

\begin{abstract}
{
The development of problem-solving competence (PSC) among high school students is foundational for preparing resilient and adaptive citizens. Generative artificial intelligence (GenAI) can support this process, but it may also encourage students to offload part of the cognitive work that is necessary for deep learning. While the individual effects of GenAI use are increasingly studied, its collective consequences for competence development within classroom environments remain underexplored. In this study, we use an agent-based model to simulate the evolution of PSC in a high school physics classroom, where students complete tasks individually, in collaboration with peers, or with the support of GenAI. By comparing classrooms with and without access to GenAI across different peer-network structures, we show that GenAI use can diminish competence development and increase the share of students remaining in lower competence tiers. These results suggest that the educational impact of GenAI should be assessed not only through individual learning outcomes but also through its effects on collective competence dynamics.
}

\end{abstract}




\section{Introduction}
{
The use of generative artificial intelligence (GenAI) among students is increasingly reshaping education, particularly with regard to the development of various competences~\cite{miao2023guidance,Qian_TechTrends_2025}. 
Competence, commonly defined as the combination of knowledge, skills, and attitudes, lies at the core of the competence-based education adopted across European educational systems and beyond. Among these competences, problem-solving competence (PSC) remains fundamental for helping students become resilient and critical thinkers in an increasingly complex and uncertain world. 
To understand how PSC develops in contemporary environments, the present study adopts a student-centred approach, closely aligned with the social constructivist educational theory~\cite{vygotsky1978great}, which positions learning as a process of knowledge construction scaffolded by social interactions. 
Drawing on this theoretical perspective, we presume that students develop PSC through either individual work or peer collaboration. 
Such a collaborative learning environment supports competence development through the co-construction of knowledge, benefiting both more knowledgeable students and those learning from them. For example, peer interactions can help sustain learners’ motivation \cite{jarvenoja2025} as well as encourage them learn from one another through peer pedagogy \cite{ching2008}.
Another potential contributor to competence development is the use of GenAI. The increasing use of GenAI among students today therefore calls for a re-examination of how problem-solving competence evolves. When used critically, GenAI may become another means of facilitating co-constructive learning, thus introducing new educational patterns. At the same time, uncritical use of GenAI may hinder competence development.
}

{
A growing body of literature suggests increasing use of GenAI among high school students and its impact on their development of PSC, particularly in the education of physics~\cite{tuveri2026role} and mathematics~\cite{vandoc2023teaching}. 
On the one hand, GenAI can support students in solving complex problems, enable individualised learning pathways~\cite{abosaooda2025evaluating}, and provide a valuable learning support mechanism~\cite{klarin2024adolescents}.
}
{On the other hand, there are significant risks regarding competence development associated with excessive dependence on GenAI systems, such as cognitive offloading~\cite{gerlich2025tools}, negative effects on critical thinking, analytical reasoning, decision-making abilities~\cite{zhai2024effects}, and eventually overall dependence on GenAI in learning contexts~\cite{bastani2025generative}. 
These raise serious concerns among educators as to how the use of GenAI among students affects their strategies for carrying out educational tasks, specifically in physics and mathematics, which are foundational for acquiring PSC.
}

{
Several studies have examined how GenAI affects students’ learning at the individual level~\cite{klarin2024adolescents,sok2025investigating,zhu2024embracing,bastani2025generative}, including students' attitudes toward GenAI and their likelihood of using it in everyday learning activities~\cite{sok2025investigating,zhu2024embracing}. 
However, considerably less attention has been paid to the classroom as a community of learners, where students’ learning strategies may influence one another. 
In other words, little is known about how different distributions of learning choices: to use or not to use GenAI, work in collaboration with peers or individually may influence the development of students’ competence, such as problem-solving, within a single classroom. 
To address this gap, this study adopts the perspective of the classroom level, investigating how students’ PSC evolves under different probabilities of undertaking learning strategies within fixed classroom conditions and what scenarios emerge when these parameters change.
These dynamics are difficult to isolate experimentally, because they depend on counterfactual combinations of GenAI access, peer collaboration, and repeated learning trajectories that are hard to control in real classrooms.
}

{
Agent-based modelling (ABM) offers a suitable methodological framework for studying learning dynamics in classrooms.
Early modelling work described teaching-learning processes through concepts from statistical physics and social interaction, showing how individual knowledge trajectories may depend on teacher influence, peer collaboration, group composition, and access to external sources of information~\cite{bordogna2001theoretical}.
Subsequent ABM studies further developed this perspective by modelling students as agents who accumulate knowledge through lessons, study, and group work, making it possible to examine temporal changes in the distribution of knowledge and grades, and the effectiveness of different teaching strategies~\cite{ormazabal2021agent}.
Other contributions have emphasised the sociocognitive dimension of learning, showing how peer-to-peer comparison and interaction with more competent peers can shape the acquisition of scientific or tiered knowledge structures~\cite{koponen2019agent}.
Taken together, this literature shows that ABM is especially useful for exploring how micro-level learning mechanisms generate classroom-level distributions of competence. 
However, ABM has not yet been substantially used to examine how students' strategic use of GenAI may reshape these distributions, for instance by compressing, widening, or otherwise reorganising the development of PSC within a classroom community.
}

{
In this work, we develop an ABM to simulate the evolution of PSC in a high school physics classroom. Students are initialised using empirical competence distributions derived from Italian educational data and are repeatedly exposed to physics exercises whose mathematical, physical, and cognitive problem-solving skill requirements are explicitly scored. At each step, students choose whether to work without GenAI support or to rely on GenAI. When they do not use GenAI, they may solve the exercise independently or interact with peers through a classroom network. Competence then evolves according to the chosen strategy, the exercise outcome, and the potential benefit of peer collaboration. By comparing classrooms with and without access to GenAI across different network structures, we examine how GenAI availability changes both average competence growth and the final distribution of competence within the classroom. Our results show that access to GenAI generally reduces the overall improvement in PSC compared with the case in which GenAI is not used. We further analyse individual trajectories to identify which students populate different parts of the final distribution. Our results suggest that the impact of GenAI on learning should be understood not only in terms of individual outcomes, but also in terms of how it reshapes collective competence dynamics within learning communities.
}

\section{Methods}

This section provides the data used to calibrate the model and the agent-based model description. 

\subsection{Data}
{
\begin{figure}[htbp]
    \centering
    \includegraphics[width=0.8\textwidth]{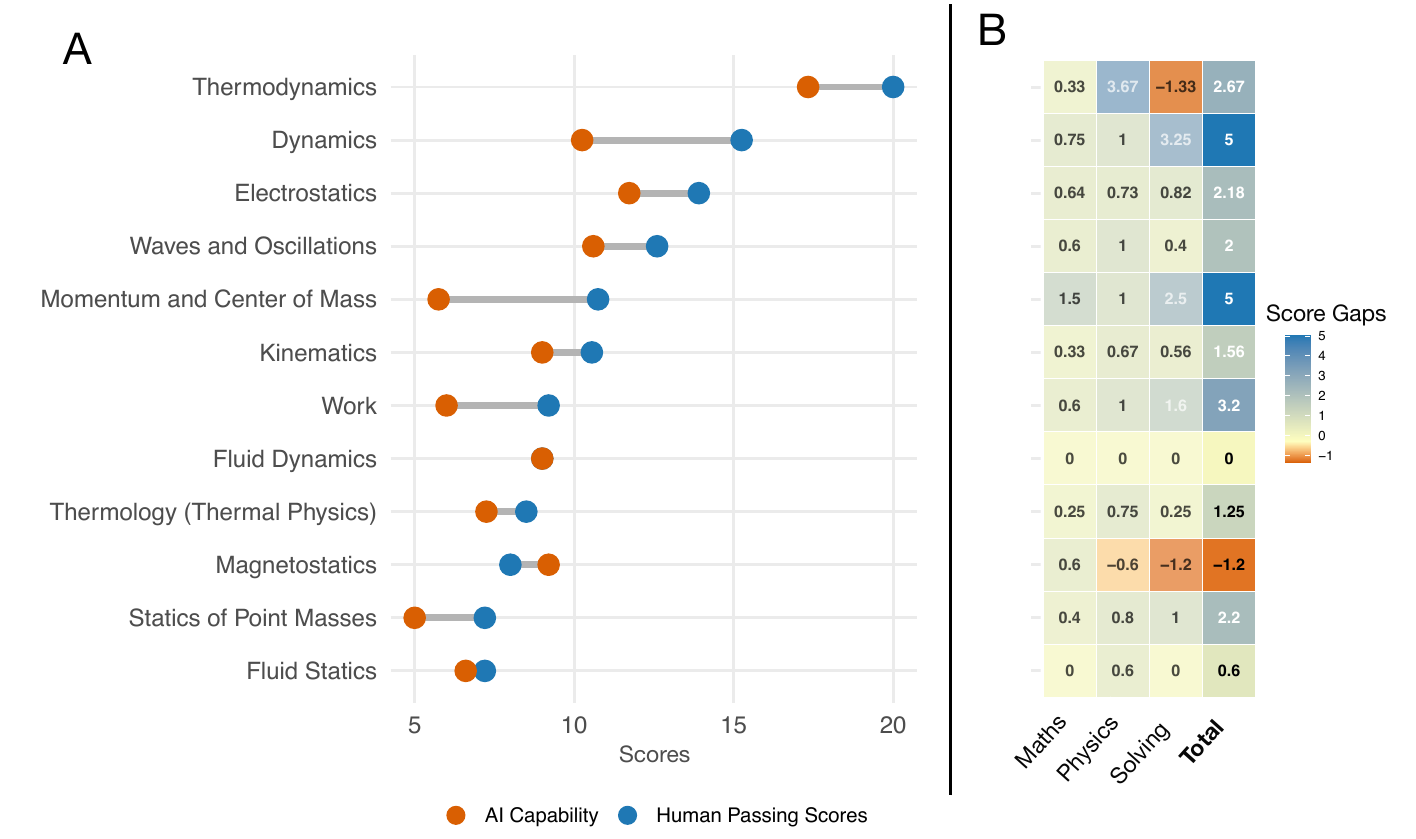}
    \caption{\textbf{Performance comparison between passing student and GenAI scores:} \textbf{(A)} Mean scores by exercise group between students (blue), that pass the exercise, and GenAI responses (orange). \textbf{(B)} Breakdown of the scoring difference between student and GenAI responses by exercise group and skills required (i.e., \textit{Maths}, \textit{Physics}, \textit{problem-solving }, \textit{Total}.)}
    \label{fig:data_exploration}
\end{figure}
}
{
We use different data sources to both initialise our model and inform the way agents acquire competences by effectively solving exercises. First, we use the INVALSI results to set the initial distribution of competences among students \cite{INVALSI2025}. INVALSI, the Italian National Institute for the Evaluation of the Education and Training System, administers national standardised assessments to monitor students’ learning outcomes in the Italian school system. These tests provide comparable information on students’ competences, primarily in Italian, English, and mathematics across different school grades and contexts. 
Second, we use a dataset of exercises at a high school level to inform how agents gain competence. This dataset consists of 60 physics exercises covering all the topics included in Italian Scientific Lyceum curricula. These exercises were extracted from a booklet \cite{virdis_esercizi} where resolutions were also provided. 

The exercises are an annotated set of physics problems grouped by topic (e.g., kinematics, electrostatics, thermodynamics). To systematically characterise the cognitive and disciplinary demands of each exercise, we developed a competence rubric spanning three domains: mathematical knowledge, physical reasoning, and cognitive problem-solving skill (CPSS). The mathematical domain encompasses sub-knowledge such as inverse formulas (1st-degree equations), systems of linear equations, square roots (2nd-degree equations), the quadratic formula, basic and advanced trigonometric functions and vector operations. At the same time, the physics domain captures two distinct dimensions: familiarity with fundamental physical formulas and the ability to apply physical reasoning (e.g., force analysis). The CPSS domain assesses procedural fluency, rewarding one point per correct reasoning step.
Each competence was scored independently, with point values reflecting its complexity: two points were assigned to higher-order sub-knowledge (e.g., systems of equations, advanced trigonometry, vector operations), while one point was assigned to more foundational knowledge. Notably, within the mathematical domain, co-occurring processes of equivalent cognitive level, such as multiple resolutions of linear equations or radicals, were not awarded additional points, as they reflect the same underlying knowledge. In the physics domain, formula recall was assigned one point for each correctly identified formula, while physical reasoning was assigned two points for each distinct reasoning step. CPSS was scored separately, with one point awarded for each procedural step. 
Each exercise was assessed under two distinct conditions. In the human condition, the exercise was solved without GenAI assistance, and mathematical knowledge, physical knowledge and CPSS scores were assigned by evaluating the human-produced solution according to the rubric described above. Therefore, these scores reflect the knowledge and reasoning steps that a student would be expected to acquire by working through the exercise using the instructor-provided textbook solution. In the GenAI condition, the same exercise was submitted to a GenAI model using a minimal prompt — “Solve this exercise” — together with a screenshot of the exercise. In the GenAI condition, the same exercise was submitted to ChatGPT using OpenAI’s GPT-5.5 Instant model \cite{openai2026gpt55instant}, accessed in May 2026.
This dual assessment yields, for every exercise, a human and a GenAI score on each component, thereby quantifying the difference in knowledge and skill acquisition between a self-studying student and one relying on GenAI assistance. Therefore, this quantifies how the expected acquisition of mathematics, physics, and CPSS differs between a student working through the exercise independently and one relying on GenAI assistance. Figure~\ref{fig:data_exploration}A compares the overall knowledge and skill scores assigned to the main exercise topics under the human and GenAI conditions. Figure~\ref{fig:data_exploration}B shows, for the same main topics, the score gap defined as the difference between the overall knowledge and skills scores assigned under the human and GenAI conditions, distinguishing between mathematics knowledge, physics knowledge, and CPSS.

A detailed example of exercise, its knowledge and skills assignment and the solution proposed by the GenAI are reported in Appendix~\ref{appendix:exercise}.

}

\subsection{Model Overview}
{
\begin{figure}[htbp]
    \centering
    \includegraphics[width=0.9\textwidth]{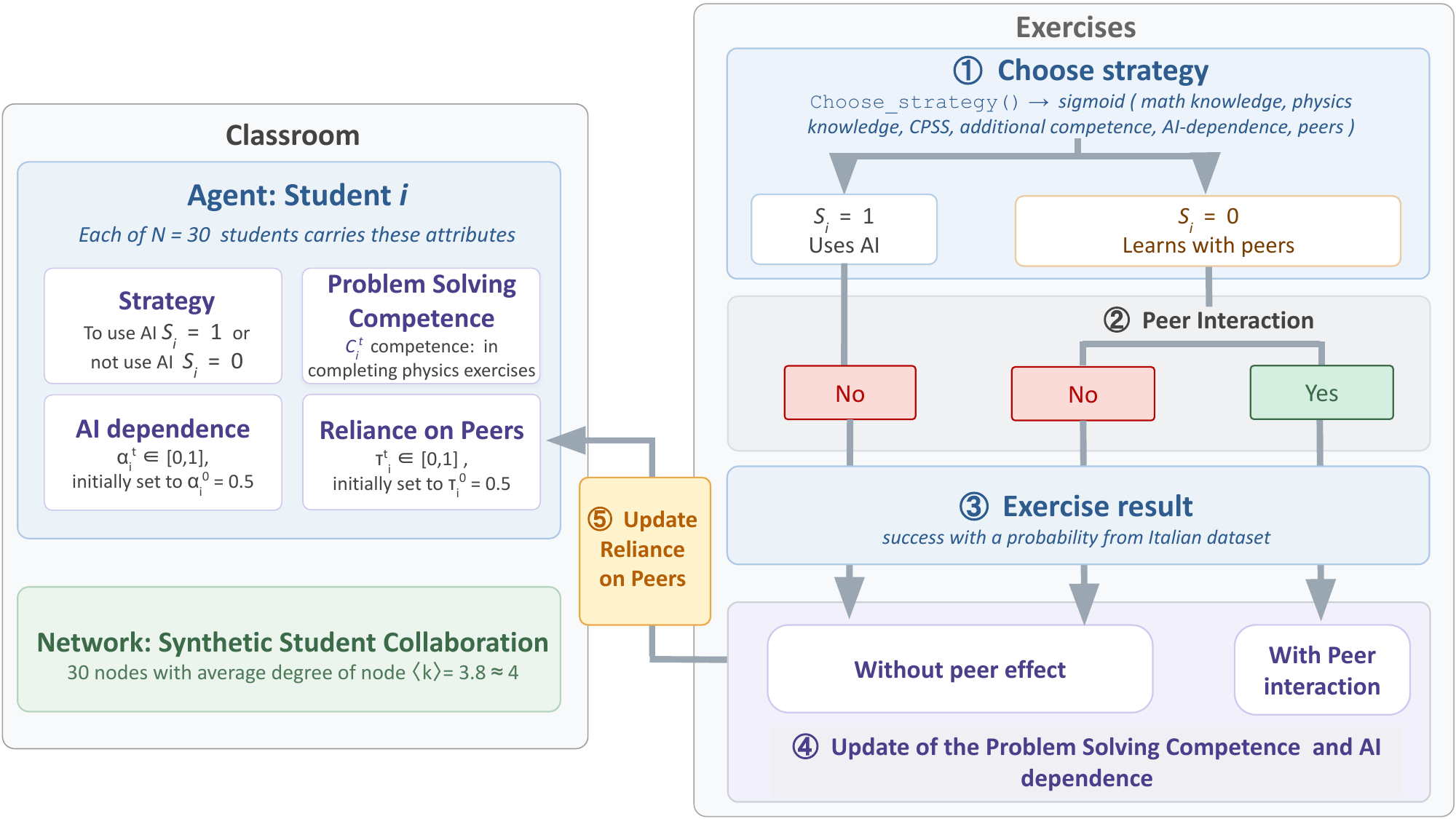}
    \caption{\textbf{Overview of the agents' dynamics and characteristics.} The diagram represents the learning process of each agent (representing a student of the classroom), connected to its peers through a network and engaged in solving physics exercises. At each time step $t$ the agent (1) choose to use or not use GenAI, (2) interact with their peers, (3) solve an exams, (4) update their problem-solving competence and GenAI dependence, and (5) update the reliance on peers.}
    \label{fig:model_vizualisation}
\end{figure}
}
{
In this study, we develop an ABM describing an ideal Italian classroom composed of $N = 30$ students. Each student is represented as an agent corresponding to a node of a graph $G$, characterised by three state variables: a problem-solving competence $C_i^t$, an GenAI dependence level $\alpha_i^t$, and a level of reliance on peers $\tau_i^t$.
These variables evolve over time $t$ by discrete steps as students repeatedly engage in a sequence of exercises. Students interact with their peers through different networks, including small-world networks, Erdős-Rényi networks, and stochastic block models.

At each discrete time step $t$, each student $i$ faces one exercise and chooses whether to use GenAI or to solve the exercise without GenAI support. If a student chooses not to use GenAI, they may either work independently or collaborate with one of their peers in the network. After this decision, the student attempts the exercise and may either succeed or fail. The student then updates their problem-solving competence according to the chosen strategy, the exercise outcome, and whether peer collaboration occurred. Finally, each student updates their level of GenAI dependence and reliance on peers. Figure \ref{fig:model_vizualisation} provides an overview of the agents’ characteristics and dynamics. In the following sections, we describe in detail each process of the model dynamics, from strategy selection to peer interaction, exercise resolution, problem solving competence update, and peer-reliance update.
}

\subsection{Agents Description and Model Dynamics}\label{sec:agent_description}
{

Each agent $i$ in the model represents a student and is characterised by three state variables: a problem-solving competence $C_i^t$, an GenAI dependence level $\alpha_i^t \in [0,1]$, and a level of reliance on peers $\tau_i^t \in [0,1]$. The problem-solving competence represents the student’s ability to solve physics exercises. The GenAI dependence level refers to how much the student relies on GenAI support, while the reliance on peers indicates the tendency of the student to seek help from classmates.
We assume no a priori distribution for the reliance on peers, so we initialise $\tau_i^0 = 0.5$ for each agent $i$. 
We also initialise the GenAI dependence $\alpha_i^0=0.5$ for each agent $i$, representing an intermediate level of GenAI dependence, consistently with recent evidence showing that approximately half of older adolescents reported using GenAI for schoolwork ($52.6\%$) \cite{klarin2024adolescents}.
The initial problem-solving competence $C_i^0$ is drawn from a normal distribution whose parameters are based on the of the 2025 INVALSI assessment \cite{INVALSI2025}; in particular, we consider the achieved levels with their corresponding percentage, and we fit a normal distribution, obtaining
$\mathcal{N}(\mu,\sigma)$ with $\mu = 195.9$ and $\sigma = 1.28$. For each student, we also track the accumulated competence gain at time $t$, defined as
$\Delta C_i^t = C_i^t - C_i^{0}$
which measures how much the student has improved since the beginning of the simulation.

Students interact with each other through peer networks. To evaluate the role of peer-network structure, we consider different network topologies, which are described in Section \ref{sec:network}.

The simulation lasts $T=60$ time steps. Each time step $t$ corresponds to the attempt to solve one physics exercise from the exercise sequence. Therefore, during the simulation, students repeatedly face exercises and update their state according to their choices and outcomes.
At each time step $t$, each student $i$ can choose between two possible strategies:

\begin{equation}
s_i^t =
\begin{cases}
0, & \text{if agent } i \text{ does not use GenAI},\\
1, & \text{if agent } i \text{ uses GenAI}.
\end{cases}
\end{equation}
When agents choose not to use GenAI, they can either work independently or collaborate with one of their peers in the network. Figure \ref{fig:model_infographics} illustrates the state of the classroom at two consecutive time steps. At each time step, students may use GenAI on their own, study alone, or cooperate with peers without GenAI. Dashed links represent the underlying network, while solid links indicate active peer interactions occurring during that time step.

\begin{figure}[htbp]
    \centering
    \includegraphics[width=0.7\textwidth]{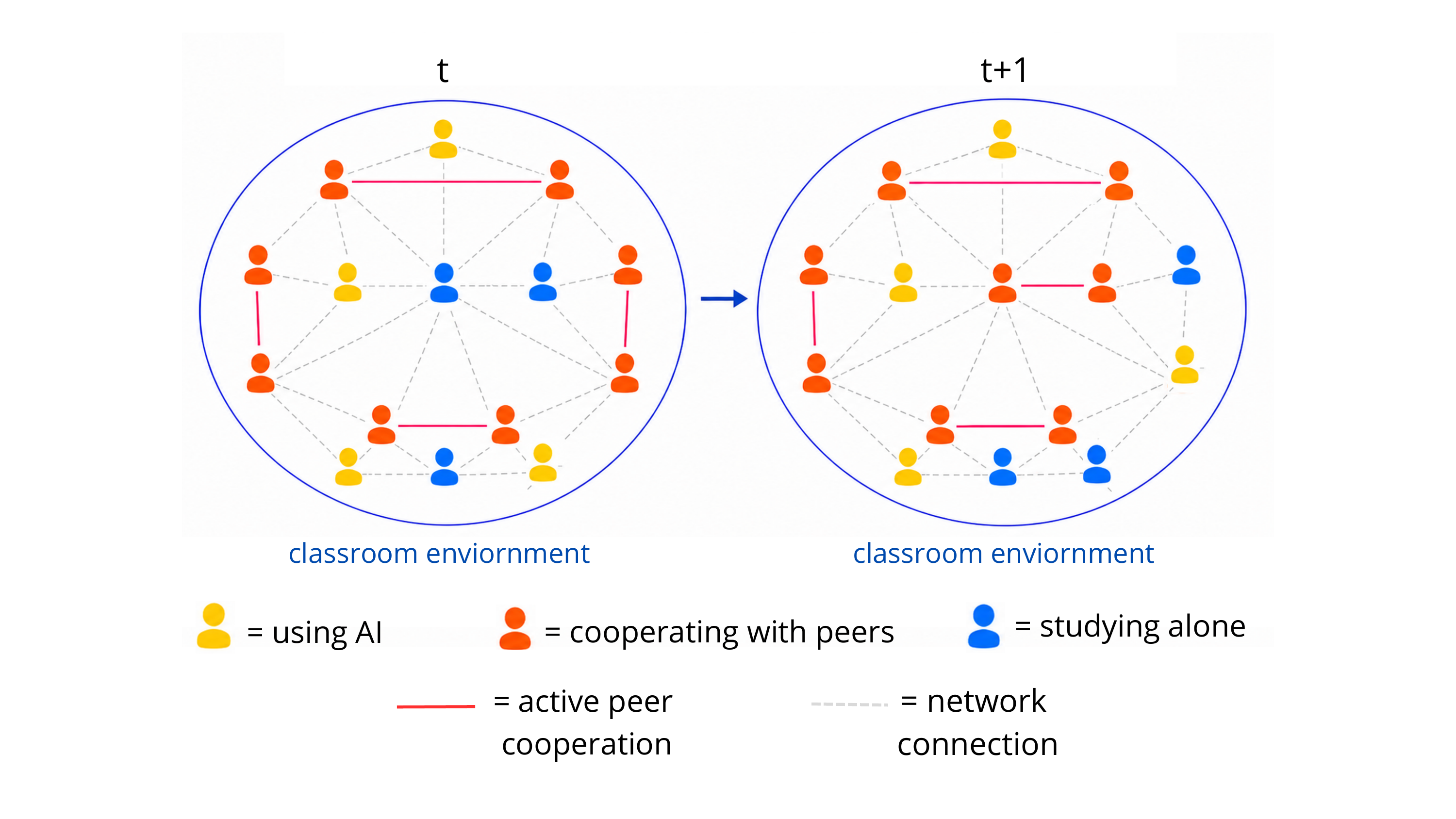}
    \caption{\textbf{Classroom state at two consecutive time steps}. Yellow agents are students using GenAI, while red and blue agents are students not using GenAI, respectively cooperating with peers and studying alone. Dashed links represent the underlying network, while red links indicate active peer collaborations.}
    \label{fig:model_infographics}
\end{figure}

At each time step in the dynamics, the agents first choose their strategies $s_i^t$, deciding whether to use GenAI or not.
Since each time step corresponds to the attempt to solve one physics exercise, at every time step $t$ the model accesses the corresponding exercise from the input dataset. Each exercise is described by six normalized components. The first three components, $k_m^t$, $k_p^t$, and $p_s^t$, represent respectively the mathematics knowledge, physics knowledge, and cognitive problem-solving skills (CPS) associated with the exercise. These quantities determine the competence gain that can be acquired when the exercise is solved without GenAI. The other three components, $k_{m,\mathrm{AI}}^t$, $k_{p,\mathrm{AI}}^t$, and $p_{s,\mathrm{AI}}^t$, describe the corresponding GenAI-related components used when the student attempts the exercise with GenAI support.

At time step $t$, the probability that student $i$ chooses to use GenAI is defined through a sigmoid function:

\begin{equation}
P(s_i^t = 1) =
\frac{1}{1+\exp\left[-\left(
w_C \left(k_m^t+k_p^t+p_s^t-\Delta C_i^t\right)
+ w_{\alpha}\alpha_i^t
- w_{\tau}\tau_i^t
\right)\right]}.
\end{equation}

where $w_C$, $w_{\alpha}$, and $w_{\tau}$ are the weights associated with the total exercise requirements, GenAI dependence, and reliance on peers, respectively. In our simulations, these weights are all set to $1/3$ to assign the same importance to the three factors entering the strategy choice.
The term $k_m^t+k_p^t+p_s^t$ represents the overall requirement of the exercise at time $t$. This term increases the probability of using AI: the more demanding the exercise, the more likely the student is to rely on GenAI support. The accumulated competence gain $\Delta C_i^t$ has the opposite effect: students who have already improved more are less likely to use GenAI. The GenAI dependence $\alpha_i^t$ increases the probability of choosing AI, while the reliance on peers $\tau_i^t$ decreases it, since students who are more willing to rely on classmates are less likely to rely on AI.
The agent will then use GenAI with probability $P(s_i^t=1)$, otherwise the student attempts the exercise without GenAI support.

As a second stage, students who have chosen not to use AI, i.e., those with $s_i^t=0$, decide whether to work independently or collaborate with one of their peers. This decision depends on the student’s reliance on peers $\tau_i^t$ in such a way that the higher is its value, the higher the probability that the student chooses to collaborate rather than work alone.
If student $i$ decides to cooperate, one of their neighbors in the peer network is randomly selected as the collaborator. 

As a third stage, all students attempt to solve the exercise associated with the current time step. The exercise outcome is modeled probabilistically and depends on the strategy previously selected by the student. In particular, for students who do not use AI, the probability of successfully solving the exercise is set to 0.54. This value is based on the 2025 INVALSI report, according to which 54\% of students involved in the analysis show adequate competences in mathematics at the national level \cite{INVALSI2025}.
For students who use AI, we assume that they always solve the exercise. This assumption reflects the idea that GenAI support allows students to obtain the solution of the exercise.

In the fourth stage of each time step $t$, students update their problem solving competence $C_i^t$ and their GenAI dependence level $\alpha_i^t$. The problem solving competence update depends on the strategy selected by the student, on the exercise outcome, and, for students not using AI, on whether peer collaboration occurred.
For students who do not use AI, i.e., $s_i^t=0$, we distinguish between successful and unsuccessful exercise attempts. If student $i$ successfully solves the exercise and works independently, its problem solving competence is updated as follows:

\begin{equation}C_i^{t+1} = C_i^t + k_m^t + k_p^t + p_s^t.\end{equation}

In this way, the student gains the full competence associated with the exercise, namely the mathematical knowledge $k_m^t$, the physical knowledge $k_p^t$, and the CPSS $p_s^t$.

If student $i$ successfully solves the exercise after cooperating with one of their random selected peers $j$, the problem solving competence update is given by

\begin{equation}
C_i^{t+1} =
\begin{cases}
\mathcal{U}(C_i^t, C_j^t) + k_m^t + k_p^t + p_s^t, & \text{if } C_j^t > C_i^t,\\
C_i^t + k_m^t + k_p^t + p_s^t, & \text{if } C_j^t \leq C_i^t.
\end{cases}
\end{equation}

where $\mathcal{U}(C_i^t, C_j^t)$ is a random value uniformly drawn between the competence of student $i$ and the competence of peer $j$. Thus, if the selected peer is more competent, student $i$ can benefit from peer collaboration. If the peer is not more competent, collaboration does not provide an additional learning effect, and the student only gains the competence associated with the exercise.

If student $i$ does not successfully solve the exercise and works independently, its problem-solving competence remains unchanged:

\begin{equation}
C_i^{t+1} = C_i^t.
\end{equation}

If, instead, student $i$ does not successfully solve the exercise but has cooperated with one of their random selected peers $j$, the competence update depends on the competence of the selected peer:

\begin{equation}
C_i^{t+1} =
\begin{cases}
\mathcal{U}(C_i^t, C_j^t), & \text{if } C_j^t > C_i^t,\\

C_i^t, & \text{if } C_j^t \leq C_i^t.
\end{cases}
\end{equation}

Thus, when the exercise is not solved, the student does not gain the competence associated with the exercise. However, if peer cooperation occurred and the selected peer is more competent, student $i$ can still improve through the interaction.

For students who choose to use AI, i.e., $s_i^t=1$, we do not distinguish between successful and unsuccessful exercise attempts in the problem solving competence update. This is because we assume that the interaction with GenAI produces an effect on the student’s $C_i^t$ regardless of the final exercise outcome. In particular, GenAI use may have a positive effect on the mathematics and physics related components of competence, while potentially reducing the development of autonomous problem-solving skills. More specifically, we assume that the maximum gain in mathematics knowledge is $k_{m,\mathrm{AI}}^t$, the maximum gain in physics knowledge is $k_{p,\mathrm{AI}}^t$, and the maximum gain in CPSS is zero. Conversely, the minimum gains are $0$ for mathematics and physics knowledge, and $-p_{s,\mathrm{AI}}^t$ for CPSS. Therefore, the problem-solving competence is updated as follows:

\begin{equation}
C_i^{t+1}
=
C_i^t
+
\mathcal{U}(0,k_{m,\mathrm{AI}}^t)
+
\mathcal{U}(0,k_{p,\mathrm{AI}}^t)
+
\mathcal{U}(-p_{s,\mathrm{AI}}^t,0).
\end{equation}

After the problem solving competence update, each student updates its GenAI dependence level. This update depends on the strategy selected by the student and on the exercise outcome. If a student does not use GenAI and successfully solves the exercise, its $\alpha_i^t$ decreases, since solving the exercise without GenAI reinforces autonomous problem solving. If a student does not use GenAI but does not successfully solve the exercise, its GenAI dependence remains unchanged. Conversely, if a student uses AI, its $\alpha_i^t$ increases, reflecting the reinforcement of dependence on GenAI support \cite{gerlich2025tools}.
More precisely, let $\eta_i^t$ be a random value uniformly drawn as follows:

\begin{equation}
\eta_i^t \sim \mathcal{U}\left(\alpha_i^t,1-\alpha_i^t
\right).
\end{equation}

The GenAI dependence level $\alpha_i^t$ update is then given by

\begin{equation}
\alpha_i^{t+1} =
\begin{cases}
\max\left(0,\alpha_i^t-\eta_i^t\right), & \text{if } s_i^t=0 \text{ and the exercise is solved},\\
\alpha_i^t, & \text{if } s_i^t=0 \text{ and the exercise is not solved},\\
\min\left(1,\alpha_i^t+\eta_i^t\right), & \text{if } s_i^t=1.
\end{cases}
\end{equation}

In the fifth and final stage of each time step, students update their level of reliance on peers. This update occurs only for students who have cooperated with one of their peers during the current time step. If no peer cooperation occurred, $\tau_i^t$ remains unchanged.
When student $i$ has cooperated with a peer, the update depends on whether the interaction has been beneficial in terms of problem-solving competence. In particular, if the student's problem-solving competence after the interaction is higher than their initial one, the level of reliance on peers increases. Conversely, if the student's problem-solving competence is lower than their initial one, the level of reliance on peers decreases.
More precisely, let $\xi_i^t$ be a
random value uniformly drawn as follows:

\begin{equation}
\xi_i^t \sim \mathcal{U}\left(\tau_i^t,1-\tau_i^t\right).
\end{equation}

The update of the reliance on peers is then given by

\begin{equation}
\tau_i^{t+1} =
\begin{cases}
\min\left(1,\tau_i^t+\xi_i^t\right), & \text{if } \Delta C_i^{t+1} > 0,\\
\max\left(0,\tau_i^t-\xi_i^t\right), & \text{if } \Delta C_i^{t+1} < 0.
\end{cases}
\end{equation}

As mentioned at the beginning of this section, agents interact through different types of networks.

Table \ref{tab:model_parameters_variables} summarizes the main parameters and variables used in the model.

\begin{table}[h!]
\centering
\caption{\textbf{Main parameters and variables used in the model.}}
\label{tab:model_parameters_variables}
\renewcommand{\arraystretch}{1.4}
\begin{tabular}{lll}
\hline
\textbf{Symbol} & \textbf{Description} & \textbf{Type} \\
\hline
$N$ & Number of students in the classroom & Model parameter \\
$T$ & Number of time steps, corresponding to the number of exercises & Model parameter \\
$\langle k \rangle$ & Average degree of the networks used to model students' interactions & Model parameter \\
$C_i^t$ & Problem-solving competence of student $i$ at time $t$ & State variable \\
$\alpha_i^t$ & GenAI dependence level of student $i$ at time $t$ & State variable \\
$\tau_i^t$ & Reliance on peers of student $i$ at time $t$ & State variable \\
$s_i^t$ & Strategy selected by student $i$ at time $t$ & State variable \\
$k_m^t$ & Mathematics knowledge component of exercise $t$ & Exercise variable \\
$k_p^t$ & Physics knowledge component of exercise $t$ & Exercise variable \\
$p_s^t$ & Cognitive problem-solving skills component of exercise $t$ & Exercise variable \\
$k_{m,\mathrm{AI}}^t$ & AI-related mathematics knowledge component of exercise $t$ & Exercise variable \\
$k_{p,\mathrm{AI}}^t$ & AI-related physics knowledge component of exercise $t$ & Exercise variable \\
$p_{s,\mathrm{AI}}^t$ & AI-related cognitive problem solving component of exercise $t$ & Exercise variable \\
\hline
\end{tabular}
\end{table}
}

\subsection{Network Structure}\label{sec:network}

We consider different network topologies to represent plausible patterns of interaction within a classroom of 30 students. The following network models are included.

{
\subsubsection{Erd\H{o}s--R\'enyi random network}

When the collaboration network is chosen as an Erd\H{o}s--R\'enyi (ER) random graph \cite{erdos1959random,gilbert1959RandomGraphs}, students have social contacts, but these contacts are assigned uniformly at random across all possible student pairs. The ER network therefore preserves only the overall interaction density and removes mesoscopic organization such as communities,  repeated-contact groups, and friendship- or availability-based constraints. 

Our ER network contains $N=30$ students and is undirected, unweighted, and without self-loops. Each possible pair of students is connected independently with the same probability. To match the empirical scale of sparse classroom interaction networks, we set the expected mean degree to $\langle k\rangle=4$, corresponding to $p_{\mathrm{ER}}\approx0.138$ and an expected total of $57$ edges. This value reflects the observation that students typically interact with only a limited number of peers rather than forming dense all-to-all collaboration networks \cite{bruun2014Time}. 

\subsubsection{Watts--Strogatz small world network}

We additionally use a Watts--Strogatz (WS) small-world network as a structural control for local clustering and shortcut-mediated reachability \cite{watts1998Collective}. Empirical collaboration networks often exhibit small-world properties, including short path lengths and high clustering in scientific and creative collaboration systems \cite{newman2001Structure,uzzi2005Collaboration}. Similar structures have also been reported in educational settings, where college-class enrolment networks form highly clustered student networks with short global distances \cite{weeden2020SmallWorld}. 

The WS network introduces local closure and short paths while keeping degree relatively homogeneous and without imposing explicit group membership \cite{amaral2000Classes,thiriot2020SmallWorld}. It therefore provides a useful comparison topology for knowledge diffusion and cooperation, where small-world networks have often been used as controlled interaction structures \cite{cowan2004Network,cassar2007Coordination}. We used the same interaction scale as in the ER layer, with $N=30, k_{\mathrm{WS}}=4$, and $60$ edges.

\subsubsection{Community-structured classroom network}

We also consider a stochastic block model (SBM) to represent structured classroom collaboration through a latent-group structure: nodes are assigned to blocks, and edge probabilities depend on block membership \cite{holland1983Stochastic,karrer2011Stochastic,peixoto2014Efficient}. 

This modelling choice is motivated by empirical studies showing that student help, problem-solving, and active-learning networks are sparse, locally structured, and organized into detectable communities or coherent subgroups \cite{bruun2014Time,vanrijsewijk2018Description,traxler2015Community,traxler2020Network,commeford2021Characterizing}. SBM-based analyses of empirical collaboration and researcher networks have also recovered interpretable group structure, further supporting the use of block-based generative models for collaboration systems \cite{ji2016Coauthorship,barbillon2017Stochastic}. We therefore divide the network into $B=5$ equal-sized blocks, giving six students per block. This empirical-scale coarse graining maps the small-to-moderate collaboration communities observed in larger student-interaction networks onto a $30$-student classroom. Each block is thus interpreted as a local collaboration group formed through friendship, availability, and repeated contact, not as an ability group.

We matched the same expected interaction scale with the ER network, $\langle k\rangle=4$, corresponding to an expected total of $60$ edges. We set $p_{\mathrm{in}}=0.47$ and $p_{\mathrm{out}}\approx 0.07$, so that ties are more likely within collaboration groups while the overall density remains matched to the ER network. Differences between the ER and SBM layers can therefore be interpreted as effects of modular collaboration structure rather than differences in the total number of peer interactions.

}
\subsubsection{Modelling competence-based homophily in the initial community structured network }\label{sbmm}

{The literature on classroom communities shows that friendship is only
one facet of peer structure, and that this structure is shaped by the
academic performance of the group \cite{qin2025}. To reproduce this community
organisation in our model, we generate the initial network with an
SBM whose blocks are defined by competence.

Each node $i$ is assigned an initial competence $C_i^0$ drawn from a
(truncated) Gaussian distribution on $[0,1]$ with mean $0.5$ and standard
deviation $\sigma$. Nodes are then partitioned into $B$ competence blocks
by thresholding $C_i^0$, and these block labels define the SBM.
An edge is placed within a block with probability $p_{\text{in}}$ and
between blocks with probability $p_{\text{out}}$, with
$p_{\text{in}} > p_{\text{out}}$ to induce homophily. We calibrate
$p_{\text{in}}$ and $p_{\text{out}}$ so that the expected mean degree
equals the target value $\langle k\rangle$ used in the unstructured baseline,
keeping network density comparable across conditions, in this case we take $p_{\text{in}}=0.70 $ and $p_{\text{out}}= 0.02$.

We assign students to blocks on the basis of initial performance, reflecting the well-established tendency of students with similar competence levels to form homophilous peer groups \cite{flashman2012}.

In particular, longitudinal network studies have shown that students tend to form and maintain peer ties with classmates of similar academic achievement, and that this form of academic homophily can persist through processes of network restructuring even when formal ability tracking is absent \cite{smirnov2017Formation}. This supports our choice to encode competence-based homophily in the initial network structure, rather than assuming an initially random peer topology.

}

\section{Results}

\subsection{Trajectories of Mean Competence across Network Topologies}
{
We consider two possible treatment conditions; in the first one agents do not have the possibility to access the GenAI, while in the second one they can access GenAI. For each combination of network structure and treatment condition, we ran a simulation considering $1000$ independent classrooms, each composed of $30$ students.

Across all four network structures analysed, the simulation results indicate a distinct variance in competence acquisition trajectories depending on the introduction of GenAI in classrooms. Figure \ref{fig:competence_trajectory_split} illustrates the mean trajectories of student competences ($C$) alongside their standard deviations ($\pm1\sigma$), derived from the ensemble of simulated paths tracked. In all considered topologies, Erdős–Rényi (ER), Stochastic Block Model (SBM), modified SBM, and Watts-Strogatz (WS), students operating without GenAI (represented by the blue solid line) consistently demonstrated a faster rate of skill acquisition compared to those in environments where GenAI is utilized as an exercise-solving tool (represented by the orange solid line in Figure~\ref{fig:competence_trajectory_split}). In aggregate, access to GenAI leads to a final state where the population of students reaches a lower average PSC as compared to the scenario without GenAI access.

}
{
\begin{figure}[htbp]
    \centering
    \includegraphics[width=0.8\textwidth]{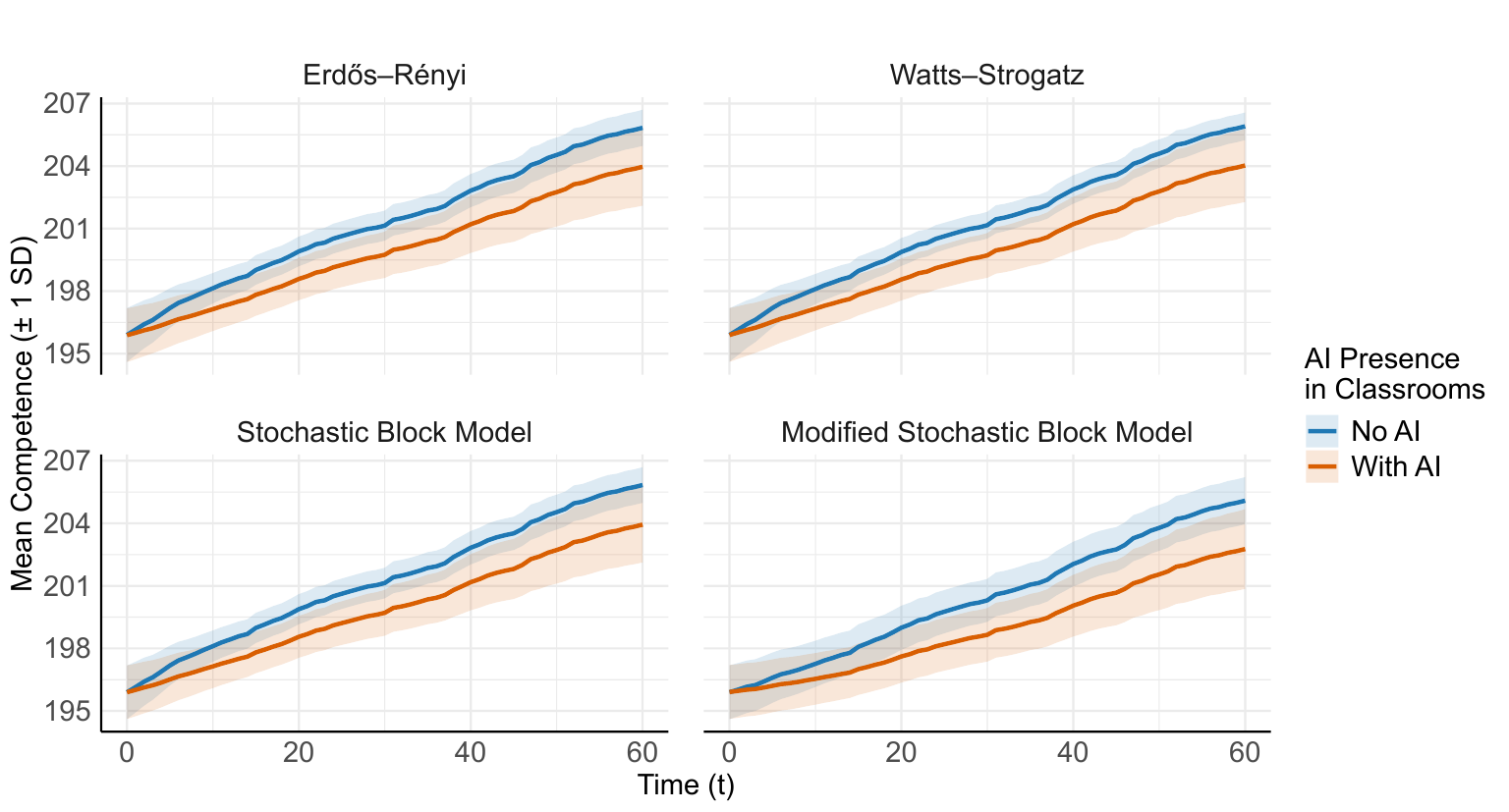}
    \caption{\textbf{Student Competency Trajectories:} Plots showing the trajectories of the mean student competency whether GenAI is present in the classrooms (blue) or not present (orange) as strategy for students  by all four network types.}
    \label{fig:competence_trajectory_split}
\end{figure}
}

\subsection{AI access slows upward mobility across competence tiers}
After showing that students with access to GenAI achieve lower values of $C$ on average, we further examine the dynamics of competence acquisition.
To this end, we consider the distribution of $C$ at three time steps, $t=0, 30, 60$, corresponding to the beginning, middle, and end of the simulation. 
At each of those time steps, students are assigned to five competence tiers by splitting the range of values taken by $C$ into five equally spaced intervals: Very Low, Low, Mid, High, and Very High.

\begin{figure}[h]
    \centering
    \includegraphics[
        width=\textwidth]{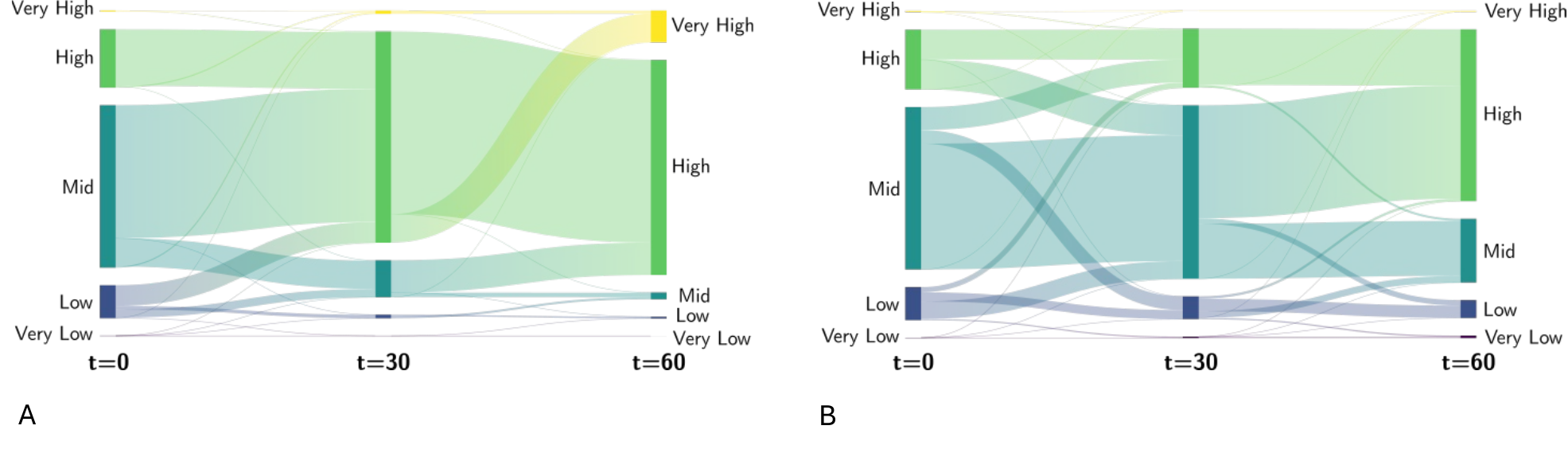}
    \caption{\textbf{Student mobility across PSC tiers.} Sankey diagrams showing transitions between five PSC tiers at $t=0, 30, 60$ in the no-GenAI (A) and GenAI (B) conditions. The tiers are obtained by splitting the PSC range into five equally spaced intervals: Very Low, Low, Mid, High, and Very High.
    This refers to the modified SBM interaction network.
    }
    \label{fig:sankey_plots}
\end{figure}

Figure~\ref{fig:sankey_plots} shows how students move across these tiers with (Figure~\ref{fig:sankey_plots}A) and without (Figure~\ref{fig:sankey_plots}B) access to GenAI on a modified SBM topology. 
In both settings, the distribution of $C$ at $t=0$ reflects the normal distribution used to initialize the $C$ within classrooms. 
The simulation dynamics, however, shows marked differences between the two treatment conditions.
Without access to GenAI, students progressively move towards higher tiers of competence, with a reduction in the number of students located in the Mid tier or below. 
By the end of the simulation, almost all students are concentrated in the High and Very High tiers.
In contrast, when students have access to GenAI, the distribution across tiers remains more stable during the first half of the simulation and displays a clear distinction from the no-AI condition by $t=60$. 
In particular, students tend to remain in the Mid tier across the simulation, with non-negligible sizes of the Low and Very Low tiers, whereas these tiers are almost entirely empty in the no-AI condition. 
Thus, while no access to GenAI leads to a relatively narrow final distribution concentrated in the upper competence tiers, having access to GenAI maintains a larger lower-competence segment of the classroom.

This is particularly marked in the competence-sorted SBM shown in Figure~\ref{fig:sankey_plots}, where peer connections are organized by initial competence. The results of the other network models are shown in Appendix~\ref{appendix:sankey}. 
Comparing to the other networks, in the modified SBM setting, having access to GenAI, Figure~\ref{fig:sankey_plots}B, almost eliminates the transition towards the Very High tier and concentrates the final population mainly in the Mid and High tiers. This indicates that GenAI access may interact with competence-based homophily by reducing the chances that students move beyond their initial neighbourhood of competence, thereby amplifying stratification rather than promoting a uniform improvement across the classroom.

These results suggest that access to GenAI does not only reduce average competence acquisition, but also changes the way students move through the competence distribution. 
More specifically, GenAI access appears to slow the upward mobility of part of the classroom, leaving a subset of students lagging behind in the acquisition process.

\subsection{GenAI access splits the classroom into two competence groups} \label{bimodal}
{
The fact that a subset of students may be left behind in the acquisition of PSC suggests that the final distribution of $C$ could display bi-modality.
Therefore, we further characterize the distribution of $C$ in the population by testing whether each population splits into distinct competence groups (PSC-groups).
To do that, we fit a one-dimensional Gaussian mixture model to each distribution, where we determine the number of  PSC-groups, $k$, by testing different values $k \in \{1, 2, \dots, 6\}$.
We select the value of $k$ minimizing the Integrated Completed Likelihood (ICL) criterion~\cite{biernacki2000assessing}, which favours well-separated clusters and yields a more conservative estimate of the number of genuinely distinct clusters.
Once the optimal $k$ is selected, each student is assigned to the most likely  PSC-group according to the fitted mixture model.

Table~\ref{tab:gmm_clusters} reports, for each network type and treatment condition, the number of identified PSC-groups together with their basic statistics.
A consistent pattern emerges across all four network structures.
With no access to GenAI, the distribution of $C$ become homogeneous, with almost all students belonging to one single  PSC-group and only a small minority, correspond to 4\% of the population at most (in the modified SBM scenario), falling into a PSC-group characterized by smaller values of $C$.
The only exception is the WS network, where all students fall into one single  PSC-group.
In contrast, when students have access to AI, the PSC-group displaying lower level of $C$ is substantially larger, comprising between $8\%$ (WS) and $12\%$ (modified SBM) of the population.
This corresponds to a roughly three- to nine-fold increase relative to the no-AI case across the different network structures.
In other words, access to GenAI does not merely shift the average PSC in the population, but it leaves a larger share of students behind in their PSC acquisition.

A second, more general effect consists of the narrowing of the PSC distribution, regardless of whether students have access to AI.
For the main PSC group (Cluster 1 in Table~\ref{tab:gmm_clusters}), the spread of the the corresponding distribution shrinks relative to the initial condition (see Sec. \ref{sec:agent_description}), with the standard deviation decreasing from $1.28$ to values between $0.62$ and $0.94$ (see also Figure~\ref{fig:competence_trajectory_split}).
This effect results from peer interactions, which enhance students' learning when they engage with peers having higher PSC.
Interestingly, the low-PSC group displays a larger standard deviation, ranging from $1.27$ to $1.67$, indicating that these students not only achieve lower PSC but are also more widely dispersed in their PSC.

\begin{table}[t!]
\centering
\caption{
\textbf{AI access splits the classroom into two competence groups.}
Gaussian mixture decomposition of the population competence distributions for each network type and treatment. 
For each cluster $k$ we report the fraction of individuals it contains, together with the mean and standard deviation of its competence. 
Dashes denote a single-mode distribution.
}
\label{tab:gmm_clusters}
\begin{tabular}[h]{llc rrr rrr}
\toprule
& & & \multicolumn{3}{c}{Cluster 0} & \multicolumn{3}{c}{Cluster 1} \\
\cmidrule(lr){4-6} \cmidrule(lr){7-9}
Network & Treatment & $k$ & Frac. & Mean & Std. & Frac. & Mean & Std. \\
\midrule
\multirow{2}{*}{ER}
 & NoAI   & 2 & 0.01 & 200.91 & 1.42 & 0.99 & 205.90 & 0.62 \\
 & withAI & 2 & 0.09 & 198.84 & 1.45 & 0.91 & 204.50 & 0.77 \\
\midrule
\multirow{2}{*}{WS}
 & NoAI   & 1 & --   & --     & --   & 1.00 & 205.91 & 0.66 \\
 & withAI & 2 & 0.08 & 198.79 & 1.40 & 0.92 & 204.46 & 0.83 \\
\midrule
\multirow{2}{*}{SBM}
 & NoAI   & 2 & 0.01 & 200.84 & 1.27 & 0.99 & 205.89 & 0.68 \\
 & withAI & 2 & 0.09 & 198.90 & 1.42 & 0.91 & 204.44 & 0.80 \\
\midrule
\multirow{2}{*}{SBM (mod.)}
 & NoAI   & 2 & 0.04 & 201.16 & 1.67 & 0.96 & 205.25 & 0.72 \\
 & withAI & 2 & 0.12 & 198.41 & 1.61 & 0.88 & 203.36 & 0.94 \\
\bottomrule
\end{tabular}
\end{table}
}

\subsection{Frequent GenAI use concentrates students in the emerging lower-PSC mode}
{
%
{
To further investigate the mechanisms driving this divergence, we analyse the evolution of the PSC distribution across the classroom by partitioning students into groups based on their frequency of GenAI usage. We examine the competence histograms at four distinct cross-sections of the simulation, namely $t=1, 20, 40, 60$. A clear segregation of competence levels emerges over time. Specifically, as stressed in Subsection~\ref{bimodal}, the system transitions from an initially symmetric, Gaussian-like distribution at the beginning of the simulation ($t=1$) into an asymmetric bimodal distribution by the end of the observed time horizon. Moreover, as depicted in Figure~\ref{fig:ai_share_group_ws_sbm_mod} the lower-competence peak is heavily populated by students with the highest frequency of GenAI usage, suggesting that heavy reliance on GenAI as a shortcut actively hinders conceptual mastery and drives population-level stratification. As emphasized in Section~\ref{bimodal}, this segregation phenomenon is highly persistent across all analysed network structures (see Appendix~\ref{appendix:evolution} for the complete set of distribution plots).
}

{
\begin{figure}[htbp]
    \centering
    \includegraphics[width=\textwidth]{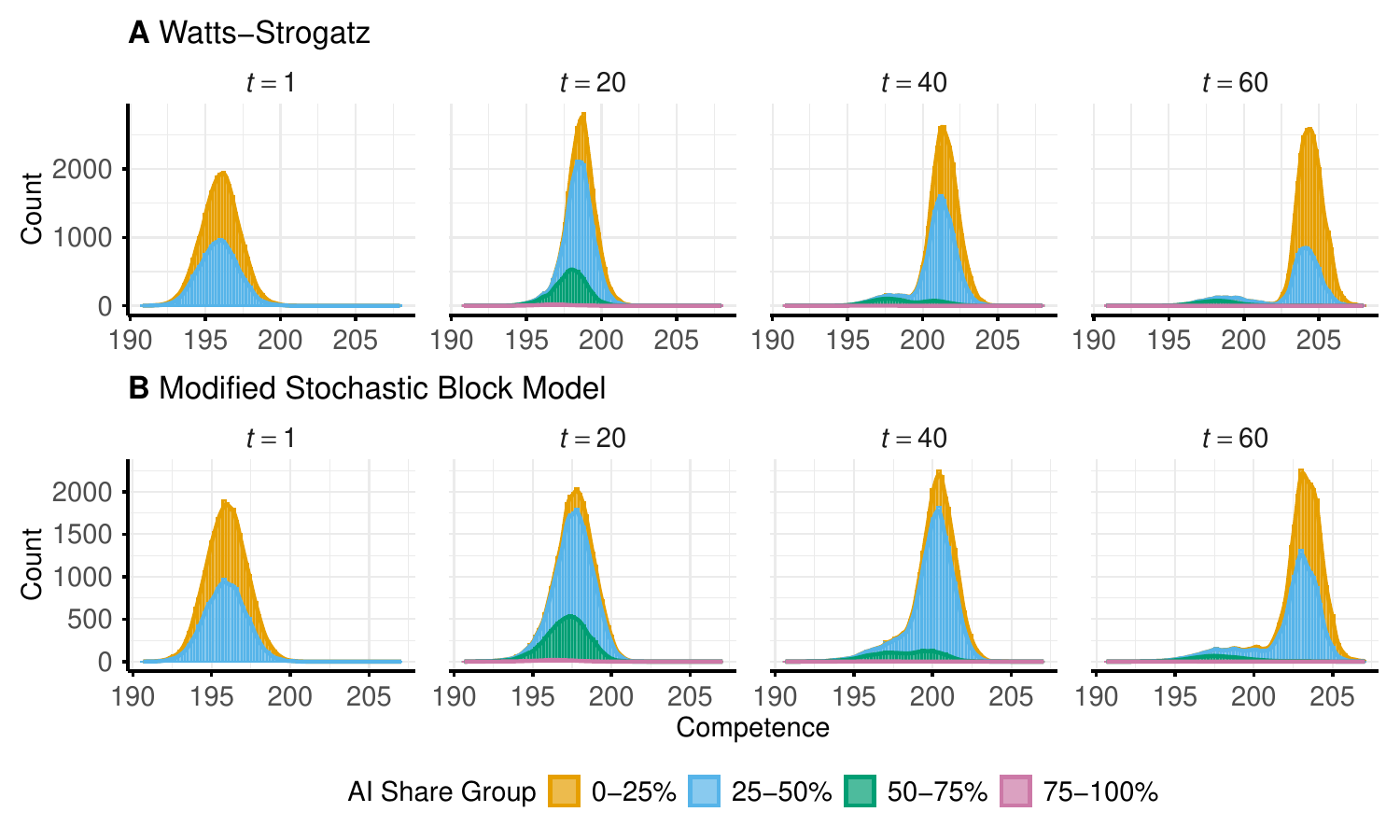}
    \caption{\textbf{Competence splits by use:} Competence histograms at $t = 1, 20, 40, 60$ for \textbf{(A)} the Watts-Strogatz network and \textbf{(B)} the modified stochastic block model, with students grouped by GenAI-usage frequency (0-100\%). A single Gaussian-like peak separates into two over time, with the heaviest GenAI users massing in the low-competence peak with heavy GenAI reliance suppressing mastery and drives stratification. This behaviour is robust across networks (see Appendix~\ref{appendix:evolution}).}
    \label{fig:ai_share_group_ws_sbm_mod}
\end{figure}
}

{
Consequently, when given the opportunity to utilize GenAI for exercise completion, total dependence does not emerge as the prevalent behavioural pattern among more proficient students; after an initial period of adoption, these advanced learners systematically decrease their relative frequency of GenAI usage.
}
}

\section{Discussion and conclusion}

The ABM developed in this study provides a framework for exploring how different patterns of the use of GenAI and peer collaboration may shape the development of PSC in a high school physics classroom. By combining individual decision rules, competence gains, and local peer interactions, the model allows us to examine not only individual learning trajectories, but also the collective competence distributions that emerge from repeated classroom-level interactions.

Overall, the simulations suggest that extensive reliance on GenAI may limit the development of PSC. Students who use GenAI for a large fraction of physics tasks tend to develop lower levels of PSC than students who do not use GenAI, or who use it only occasionally. In particular, students who rely on GenAI for at least 75\% of the tasks lag behind peers who either avoid GenAI or use it rarely (up to 25\% of their time). This pattern indicates that, under the assumptions of the model, frequent AI-assisted completion of exercises may reduce opportunities to acquire the mathematical, physical, and procedural components of PSC.

The results also point to an indirect mechanism mediated by peer interaction. Students who do not use GenAI interact more frequently with their peers and, in the simulations, tend to reach higher levels of PSC. By contrast, students who rely more heavily on GenAI are less exposed to peer learning, which reduces the positive effect of collaboration on competence development. This produces not only lower average competence among frequent GenAI users, but also greater heterogeneity in their final outcomes.

The role of peer structure is particularly visible in the modified SBM setting, where students interact preferentially with peers of similar initial competence. In this case, students who have an initial very high level of PSC are more likely to reach high levels of PSC, while only a small fraction remains at low competence levels (see Figure~\ref{fig:sankey_plots}B). Conversely, among students starting with very low PSC level, none reaches the highest tier. Moreover a proportion of students starting from middle and low levels get in low or very low competence tiers. These results suggest that organising peers into groups based on competence can amplify the lagging-behind effect.

More broadly, the simulations show that the number of students who continue to use GenAI while maintaining low PSC gradually decreases over time. This raises an important interpretive question: whether students reduce their reliance on GenAI because their competence improves, or whether persistent low competence makes GenAI use less effective or less attractive within the model dynamics. This issue connects to a broader debate in the educational literature on students' motives for using GenAI. Students may use GenAI as a tool for deeper engagement with learning, but they may also use it for cognitive offloading and task completion with reduced effort. Distinguishing between these mechanisms is essential for understanding when GenAI supports learning and when it may instead weaken competence development.

\subsection{Limitations and future work}
While modelling educational phenomena aims to improve understanding of the processes underlying learning, many aspects of PSC development remain difficult to represent. The model should therefore be interpreted as a simplified abstraction that captures selected mechanisms of PSC development while necessarily omitting others. These limitations largely arise from the multidimensional, context-dependent, and emergent nature of PSC.

First, PSC comprises academic knowledge, cognitive problem-solving skills, and students' attitudes toward learning and their own competence development. The empirical data used in the agent-based model primarily capture the first two dimensions, while students' attitudes remain outside the scope of the current work. Consequently, although the model explores how PSC evolves under different interaction patterns, such as peer collaboration and GenAI use, it does not distinguish between the motivations underlying these strategies. For example, students may use GenAI to support their learning, but they may also use it to reduce cognitive effort through offloading or to complete tasks with minimal engagement. These different motivations are likely to affect both individual learning trajectories and emergent classroom dynamics, but they are not currently represented in the model. Incorporating measures of students' attitudes, motivations, and intentions toward GenAI would therefore strengthen the explanatory capacity of the framework.

Second, the behavioural rules governing students' decisions had to be simplified in order to represent learning behaviour within an agent-based framework. Although the model distinguishes between peer interaction and GenAI use, it does not account for several factors that may influence decision-making in real classrooms, such as self-regulation, teacher support, classroom norms, or the quality of students' social relationships. Including these factors would provide a richer representation of how PSC evolves in educational settings.

Furthermore, the model is tailored to a specific educational context: a high-school physics classroom. Its parameters and assumptions therefore reflect patterns characteristic of this setting. Since educational environments vary across subjects, age groups, study programmes, institutions, and national contexts, the model should not be interpreted as universally applicable, but rather as representative of the specific context examined in this study.

An additional limitation concerns the absence of a saturation mechanism in the current competence dynamics. In the present formulation, PSC can in principle continue to increase with the number of exercises, rather than approaching an upper competence level or a plateau. This simplification is acceptable over the relatively short horizon considered here, but it limits the interpretation of longer-term trajectories. Future versions of the model should introduce saturation effects, calibrated either from longitudinal learning data, survey-based evidence on competence development, or simulations over longer instructional sequences beyond the 60 exercises considered in this study.

Finally, although the model is informed by empirical data and relevant literature, validating the emergent patterns remains challenging. The results should therefore be understood as simulation-based hypotheses rather than direct empirical evidence. Future work should compare the model outcomes with longitudinal empirical observations from different educational settings, including repeated measures of PSC, GenAI use, peer interaction, and students' motivations. Such comparisons would help assess the robustness, explanatory power, and broader applicability of the model.

Overall, this study suggests that GenAI can reshape competence development by affecting classroom-level learning dynamics. Within the assumptions of the model, extensive reliance on GenAI tends to weaken PSC development, whereas peer-based learning supports upward mobility. These results highlight the need to integrate GenAI in ways that complement, rather than replace, students' own problem-solving practice and collaborative learning.

\section*{Acknowledgements}
This work is the output of the workshop Complexity72h by Complexity Next Gen, held at Northeastern University London, London, UK, 22-26 June 2026. www.complexitynextgen.org/complexity72h/.

\bibliographystyle{unsrtnat}
\bibliography{references}

\clearpage

\appendix
\counterwithin{figure}{section}
\counterwithin{table}{section}
\renewcommand{\thefigure}{\thesection.\arabic{figure}}
\renewcommand{\thetable}{\thesection.\arabic{table}}

\section{Example Exercise Used in the Study}
\label{appendix:exercise}
Figure~\ref{fig:example_exercise} shows a representative example of physics exercise drawn from the collection in Ref.~\cite{virdis_esercizi}. The right panel graphically displays the differences between the mathematical, $k_m$, physical skills, $k_p$ and CPSS $p_s$ scores between human and GenAI modes.
{
\begin{figure}[htbp]
    \centering
    \includegraphics[width=0.8\textwidth]{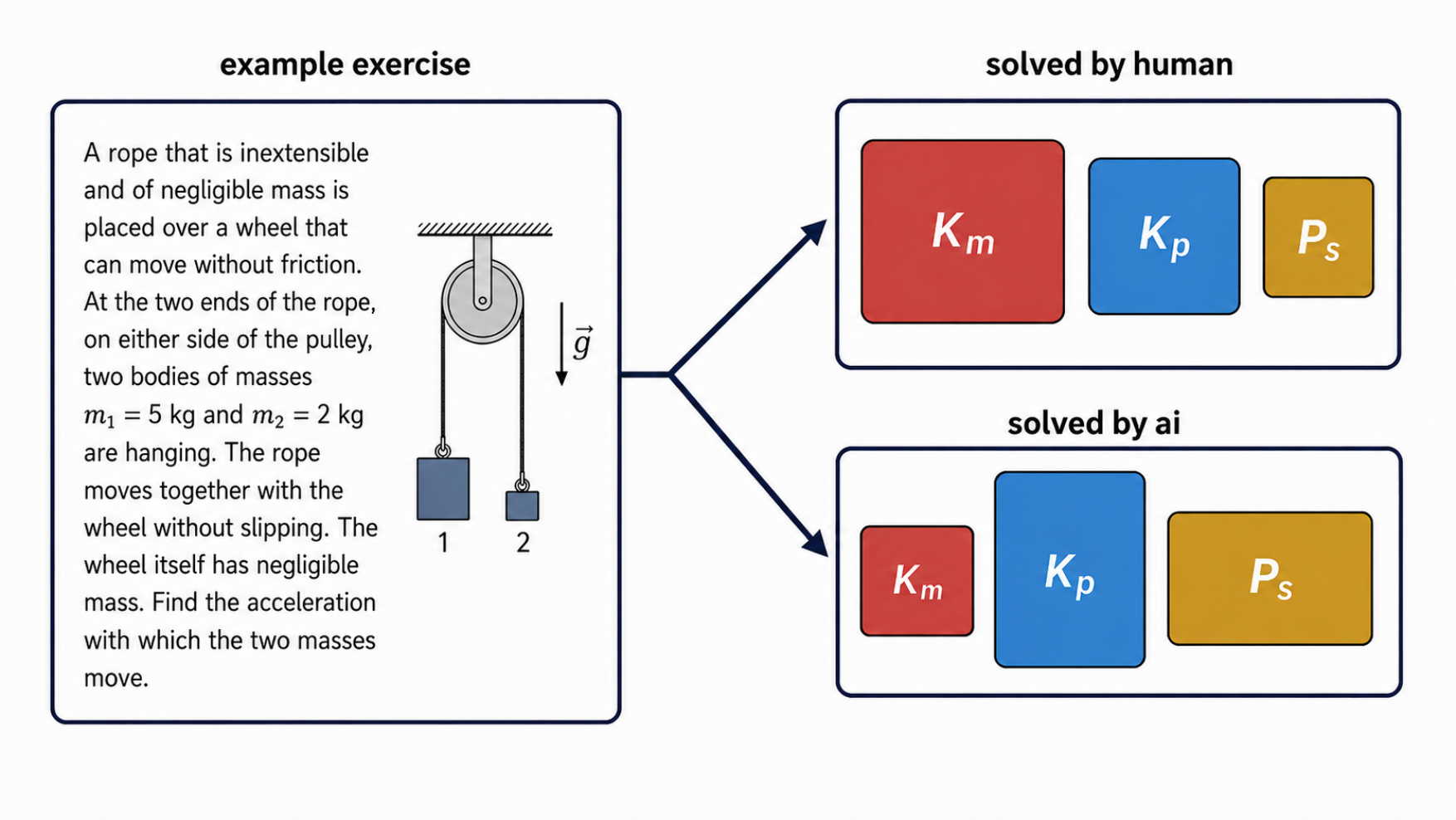}
    \caption{\textbf{Example exercise:} on the left part of the figure we reported the physics exercise, while on the right part we reported the differences in scores based on the resolution of the problem by human side and the resolution by ai.}
    \label{fig:example_exercise}
\end{figure}
}

This is the way in which ChatGPT-5.5 Instant resolved the exercise: 

\begin{solutionbox}
It is an Atwood machine with a massless pulley and a massless inextensible rope.
Since the rope does not slip and the pulley has negligible mass, the tension is
the same on both sides.

Let \(m_1 = 5\,\mathrm{kg}\) be the heavier mass and \(m_2 = 2\,\mathrm{kg}\)
the lighter mass. The heavier mass \(m_1\) moves downward, while \(m_2\) moves
upward, with the same acceleration \(a\).

For \(m_1\), taking downward as positive, we have
\[
m_1 g - T = m_1 a .
\]

For \(m_2\), taking upward as positive, we have
\[
T - m_2 g = m_2 a .
\]

Adding the two equations gives
\[
m_1 g - m_2 g = m_1 a + m_2 a ,
\]
that is,
\[
(m_1 - m_2)g = (m_1 + m_2)a .
\]

Therefore,
\[
a = \frac{m_1 - m_2}{m_1 + m_2}g .
\]

Substituting the numerical values,
\[
a = \frac{5-2}{5+2}g ,
\]
so
\[
a = \frac{3}{7}g .
\]

Using \(g \simeq 9.81\,\mathrm{m/s^2}\), we obtain
\[
a = \frac{3}{7}\cdot 9.81 \simeq 4.20\,\mathrm{m/s^2}.
\]

Thus, the acceleration is
\[
\boxed{a \simeq 4.2\,\mathrm{m/s^2}} .
\]

The \(5\,\mathrm{kg}\) mass accelerates downward, while the
\(2\,\mathrm{kg}\) mass accelerates upward.
\end{solutionbox}

\section{Representative networks}

\begin{figure}[!h]
    \centering
    \includegraphics[width=0.8\textwidth]{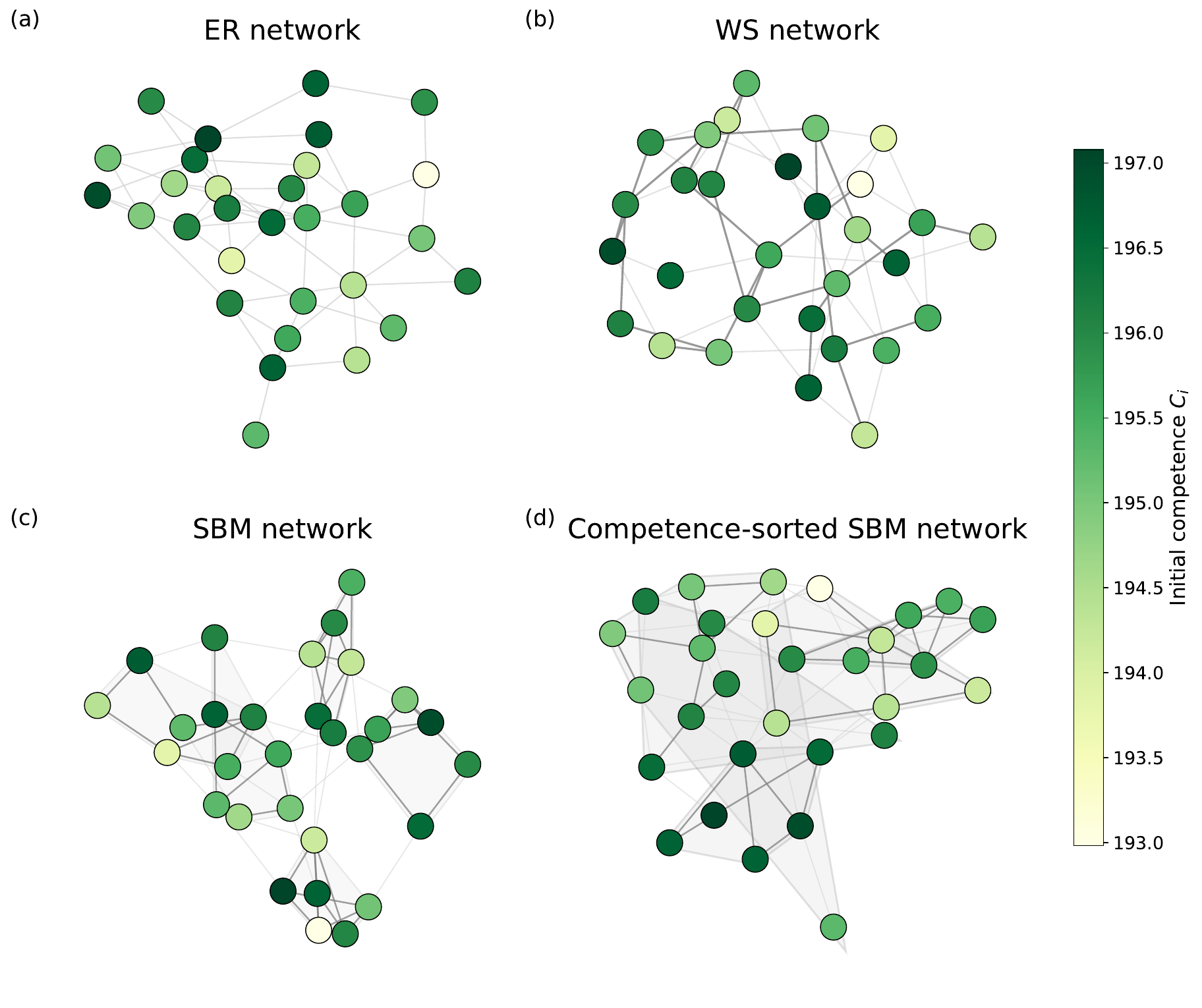}
    \caption{\textbf{Representative visualization of the four networks.} Node colour denotes initial competence $C_i(t_0)$, using a shared colour scale across all panels.
    (a) The ER network provides a density-matched random-mixing baseline.
    (b) The WS network introduces local closure and shortcut-like links while avoiding explicit group assignment.
    (c) The SBM network introduces block-structured collaboration groups; light grey hulls indicate the imposed blocks, while darker edges indicate within-block ties.
    (d) The competence-sorted SBM network uses the same block-generation process as the SBM network, but assigns nodes with similar initial competence to the same block, thereby coupling collaboration structure with competence similarity.}
    \label{fig:networks}
\end{figure}

Figure~\ref{fig:networks} provides representative visualizations of the four network layers used in the simulations. All networks contain $N=30$ nodes and use the same interaction scale, while differing in the structural feature being controlled. The ER random network represents random mixing, the WS small-world network introduces small-world local closure and shortcut-mediated reachability, the SBM introduces collaboration-group structure, and the competence-sorted SBM aligns collaboration groups with initial competence.

\newpage
\section{Students mobility across PSC tiers using different networks}
\label{appendix:sankey}
In figures \ref{fig:comparison_ai_noaiER}, \ref{fig:comparison_ai_noaiSBM}, \ref{fig:comparison_ai_noaiWS}, the comparisons between the no-GenAI and GenAI conditions in the ER, SBM and WS network model respectively are depicted. All the tiers are obtained by splitting the PSC range into five equally
spaced intervals: Very Low, Low, Mid, High, and Very High.

{

\begin{figure}[!h]
    \centering
    \begin{subfigure}{0.48\textwidth}
        \centering
        \includegraphics[width=0.9\textwidth]{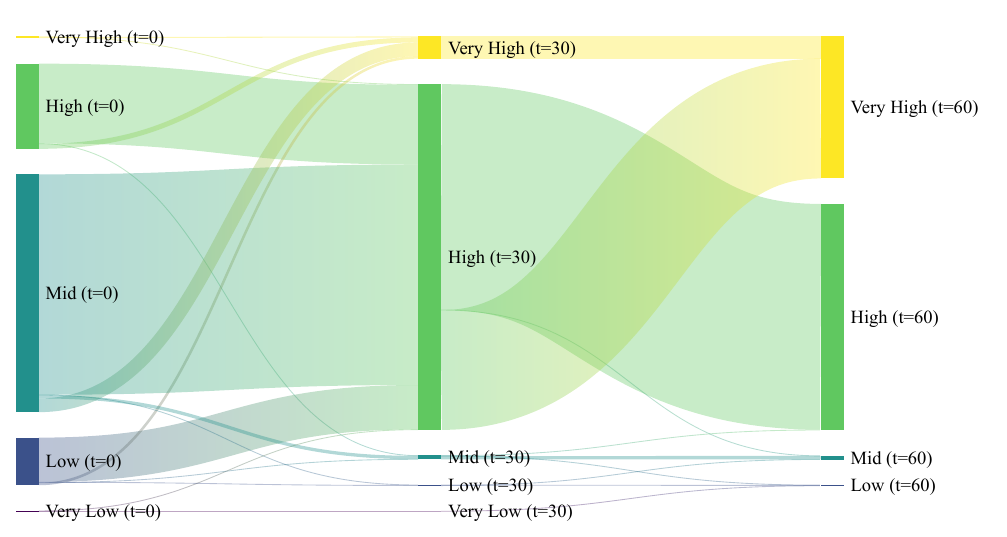}
        \caption{No-AI condition}
        \label{fig:sbm_withai}
    \end{subfigure}
    \hfill
    \begin{subfigure}{0.48\textwidth}
        \centering
        \includegraphics[width=0.9\textwidth]{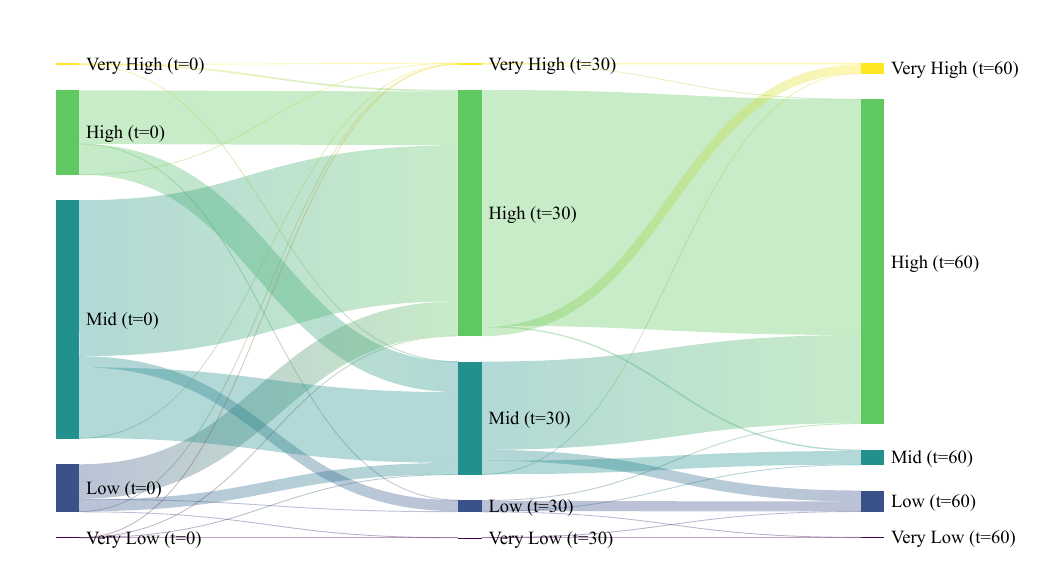}
        \caption{AI condition}
        \label{fig:sbm_noai}
    \end{subfigure}

    \caption{Sankey diagrams showing transitions between five PSC tiers at
$t = 0, 30, 60$ in the no-GenAI (A) and GenAI (B) conditions. This refers to the ER interaction network.}
    \label{fig:comparison_ai_noaiER}
    \end{figure}
}

{

\begin{figure}[!h]
    \centering
    \begin{subfigure}{0.48\textwidth}
        \centering
        \includegraphics[width=0.9\textwidth]{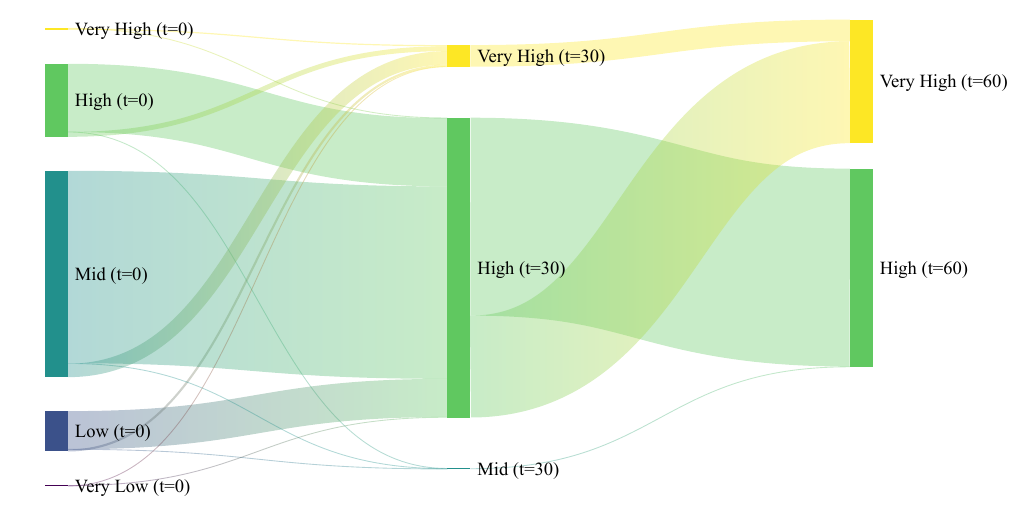}
        \caption{No-AI condition}
        \label{fig:sW_withai}
    \end{subfigure}
    \hfill
    \begin{subfigure}{0.48\textwidth}
        \centering
        \includegraphics[width=0.9\textwidth]{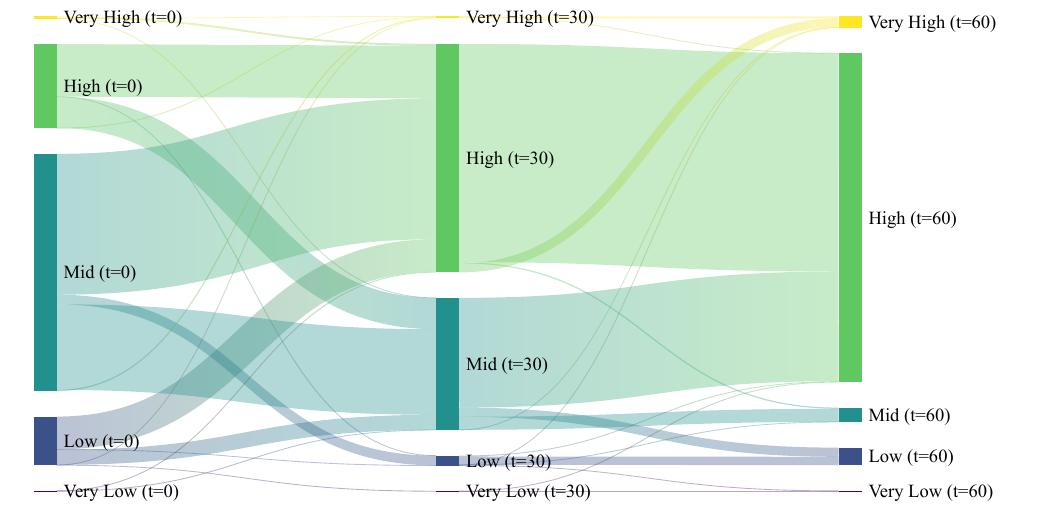}
        \caption{AI condition}
        \label{fig:sW_noai}
    \end{subfigure}

    \caption{Sankey diagrams showing transitions between five PSC tiers at
$t = 0, 30, 60$ in the no-GenAI (A) and GenAI (B) conditions. This refers to the WS interaction network.}
    \label{fig:comparison_ai_noaiWS}
    \end{figure}
}

{
\begin{figure}[!h]
    \centering
    \begin{subfigure}{0.48\textwidth}
        \centering
        \includegraphics[width=0.9\textwidth]{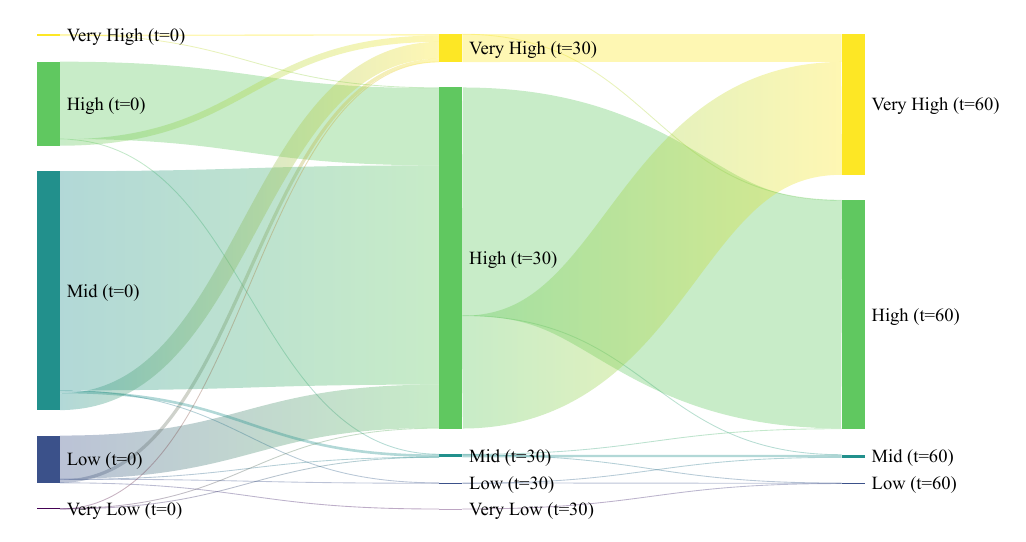 }
        \caption{No-AI condition}
        \label{fig:sbm_withai}
    \end{subfigure}
    \hfill
    \begin{subfigure}{0.48\textwidth}
        \centering
        \includegraphics[width=0.9\textwidth]{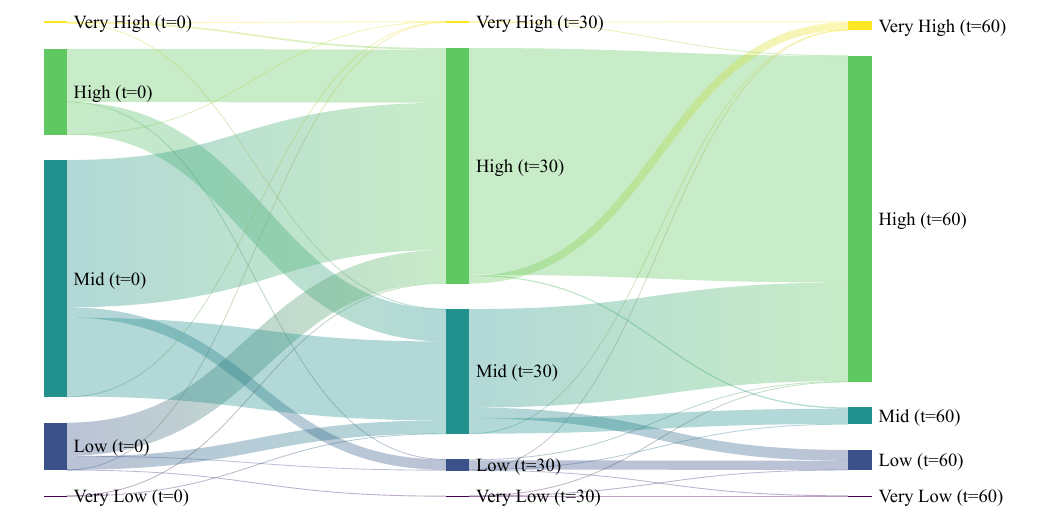}
        \caption{AI condition}
        \label{fig:sbm_noai}
    \end{subfigure}

    \caption{Sankey diagrams showing transitions between five PSC tiers at
$t = 0, 30, 60$ in the no-GenAI (A) and GenAI (B) conditions. This refers to the SBM interaction network.}
    \label{fig:comparison_ai_noaiSBM}
    \end{figure}
}

\newpage
\section{Competence Evolution across Network Types}
\label{appendix:evolution}
\subsection{No GenAI in the Classrooms}
{
Figure~\ref{fig:appendix_competence_noAI} depicts the distribution of competence levels across all analysed network structures within the baseline environment where students operate without GenAI access. As illustrated, the competence distribution mirrors a symmetric, Gaussian-like probability density. When examining the histograms across four distinct cross-sections of the simulation ($t=1,20,40,60$), the distribution behaves as anticipated: the mean systematically shifts to the right, while the peaks become increasingly narrow, reflecting a progressive homogenization of student skills. 
\begin{figure}[!h]
    \centering
    \includegraphics[width=.8\textwidth]{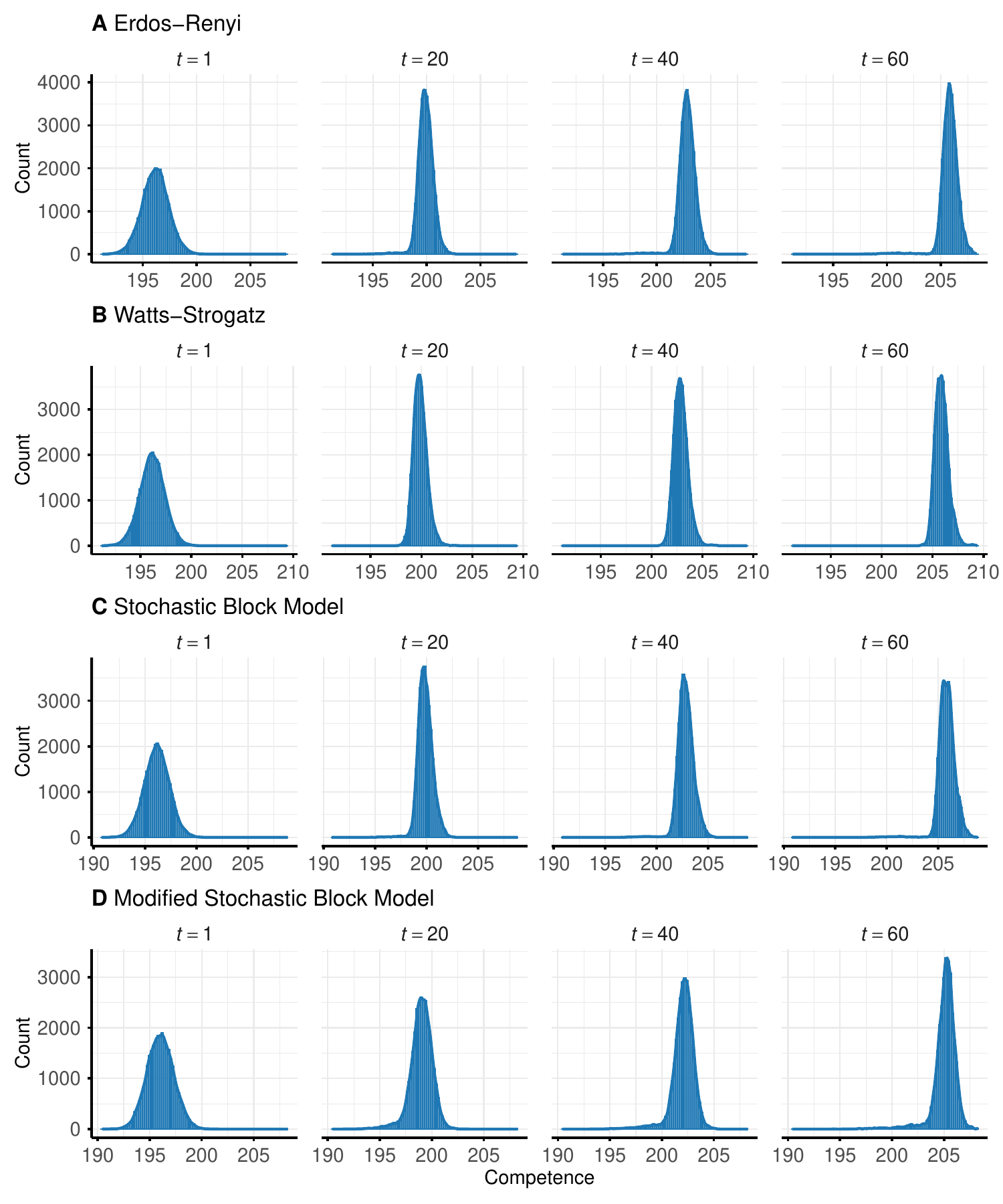}
    \caption{Distribution of PSC levels across all analysed network structures within the baseline environment where students operate without GenAI access.}
    \label{fig:appendix_competence_noAI}
\end{figure}
}

\subsection{With GenAI in the Classrooms}
{
Figure \ref{fig:appendix_competence_withAI} depicts the distribution of competence levels across all analysed network structures where students operate with the GenAI access.
As stressed in subsection \ref{bimodal}, the system transitions from an initially symmetric, Gaussian-like distribution at the baseline into an asymmetric bimodal distribution by the end of the observed time horizon. Moreover, as depicted in Figure \ref{fig:appendix_competence_withAI} the lower-competence peak is heavily populated by students with the highest frequency of GenAI usage, suggesting that heavy reliance on GenAI as a shortcut actively hinders conceptual mastery and drives population-level stratification. As emphasized in Section \ref{bimodal}, this segregation phenomenon is highly persistent across all analysed network structures (A-D).
\begin{figure}[!h]
    \centering
    \includegraphics[width=.8\textwidth]{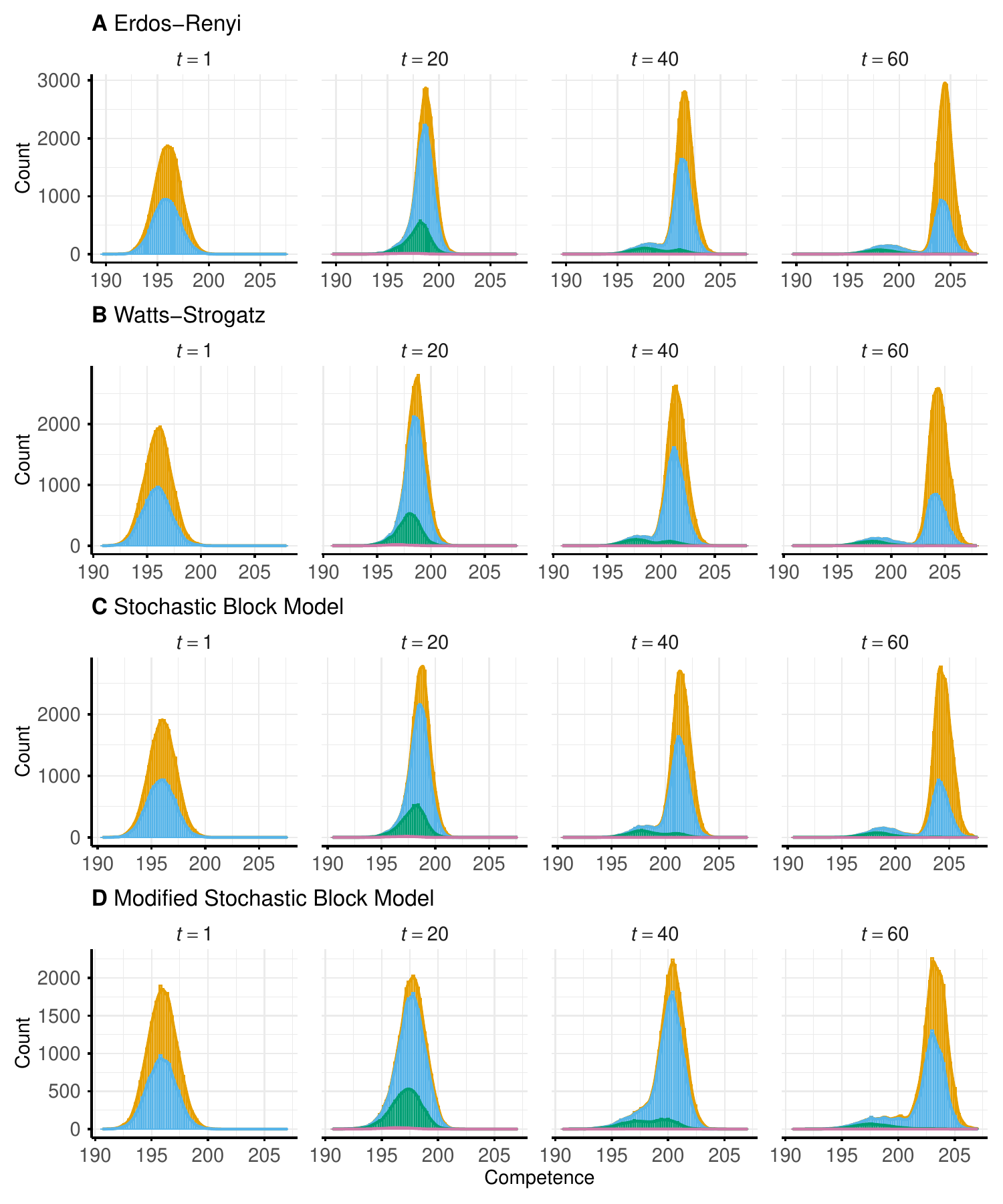}
    \caption{Distribution of PSC levels across all analysed network structures where students operate with the GenAI access.}
    \label{fig:appendix_competence_withAI}
\end{figure}
}


%


%

\end{document}